\documentclass[aps,twocolumn,a4]{revtex4}
\usepackage{epsfig}
\usepackage{psfrag}
\newcommand{\be}{\begin{equation}}
\newcommand{\ee}{\end{equation}}
\newcommand{\ba}{\begin{array}}
\newcommand{\ea}{\end{array}}

\begin{document}
\title{Structure optimization  in an off-lattice protein model}
\author{Hsiao-Ping Hsu, Vishal Mehra and Peter Grassberger}
\affiliation{John-von-Neumann Institute for Computing, Forschungszentrum
J\"ulich, D-52425 J\"ulich, Germany}

\date{\today}
\begin{abstract}
We study an off-lattice protein toy model with two species of 
monomers interacting through modified Lennard-Jones interactions.
Low energy configurations are optimized using the 
pruned-enriched-Rosenbluth method (PERM), hitherto employed to 
native state searches only for off lattice models. For 2 dimensions we
found states with lower energy than previously proposed putative 
ground states, for all chain lengths $\ge 13$. This indicates that
PERM has the potential to produce native states also for more realistic 
protein models. For $d=3$, where no published ground states exist, we
present some putative lowest energy states for future comparison with
other methods.

\end{abstract}
\maketitle

Predicting the structure of a protein, given its sequence of amino 
acids, is one of the central problems in computational biology. Since 
the problem is too difficult to be approached with fully realistic 
potentials derived from first principles, many authors
have studied it in various degrees of simplifications. This involves
in particular neglect of solvent water, simplifying the interactions, 
lumping together smaller groups of atoms, and putting everything on a 
discrete lattice. Among the most radically simplified models is the HP 
model of Dill and coworkers \cite{dill} where each amino acid is treated 
as a point particle on a regular (quadratic or cubic) lattice, and 
only two types of amino acids -- hydrophobic (H) and polar (P) -- are 
considered. Apart from the forces responsible for the connectedness of 
the chain, the only forces are contact forces between nearest lattice 
neighbours which are different for HH, HP, and PP pairs.

Even in this highly simplified model it is far from trivial to predict 
the native state for a given amino acid sequence 
\cite{yd93_95,bd96,chikenji,lw01,hmng03,backofen}.
The most efficient algorithms are either deterministic and cannot 
be generalized to more realistic models at all \cite{backofen},
or use sequential importance sampling with re-sampling in the form
of the pruned-enriched-Rosenbluth method (PERM) \cite{hmng03}. Although
it was shown that the latter can be applied also to off-lattice homopolymers
at higher temperatures \cite{g97}, it is not obvious that it will be 
efficient for off-lattice heteropolymers at the low temperatures needed 
for protein folding.

While there are a large number of benchmark cases for lattice protein models 
in the literature, there exist very few simple off-lattice models with known 
lowest energy states that can be used as benchmarks for efficient 
algorithms. One such model is the so-called AB model by Stillinger {\it 
et al} \cite{shh93,sh95} which also uses only two types of monomers, called 
now ``A" (hydrophobic) and ``B"(polar). The distances between consecutive 
monomers along the chain are held fixed to $b=1$, while non-consecutive 
monomers interact through a modified Lennard-Jones potential. In addition,
there is an energy contribution from each angle $\theta_i$ between 
successive bonds. More precisely, the total energy for a $N$ monomer chain 
is expressed as 
\be
   E = \sum_{i=2}^{N-1}E_1(\theta_i)+\sum_{i=1}^{N-2}\sum_{j=i+2}^{N}
       E_2(r_{ij},\zeta_i,\zeta_j), \label{pot}
\ee
where
\be
   E_1(\theta_i)=\frac{1}{4} (1-\cos \theta_i),
\ee
\be
   E_2(r_{ij},\zeta_i,\zeta_j)=4(r_{ij}^{-12}-C(\zeta_i,\zeta_j)r_{ij}^{-6}).
\ee
Here $r_{ij}$ is the distance between monomers $i$ and $j$ (with $i<j$).
Each $\zeta_i$ is either $A$ or $B$, and $C(\zeta_i,\zeta_j)$ is $+1,
+\frac{1}{2},$ and $-\frac{1}{2}$ respectively, for $AA,BB,$ and $AB$ pairs, 
giving strong attraction between $AA$ pairs, weak attraction between $BB$
pairs, and weak repulsion between $A$ and $B$.

This model has been studied in several papers \cite{shh93,sh95,ipp,ipp3d,gorse02}
For its 2-d version, putative ground states for various $AB$ sequences and 
for various chain lengths are given in \cite{shh93,sh95,gorse02}. 
Similar models were also studied in \cite{ipp,ipp3d,tlp01,gorse01}, but 
putative ground states for these generalizations were not given at all or 
for very short chains only.
The methods used to find low energy states of the AB model include neural
networks \cite{shh93}, conventional Metropolis type Monte Carlo procedures
\cite{sh95}, simulated tempering \cite{ipp3d}, multicanonical Monte 
Carlo \cite{ipp}, biologically motivated methods \cite{gorse01,gorse02},
and molecular dynamics \cite{tlp01}. In all cases the 
stochastic minimization can only lead to some state in the neighbourhood
of a local (and hopefully also global) minimum. A greedy deterministic 
method such as conjugate gradient descent is subsequently applied to reach
the minimum itself.

\begin{figure}[ht]
\begin{center}
$\begin{array}{c@{\hspace{0.1in}}c}
\multicolumn{1}{l}{\mbox{\tiny N = 13}} &
        \multicolumn{1}{l}{\mbox{\tiny N = 21}} \\ [-0.53cm]\\
\epsfig{file=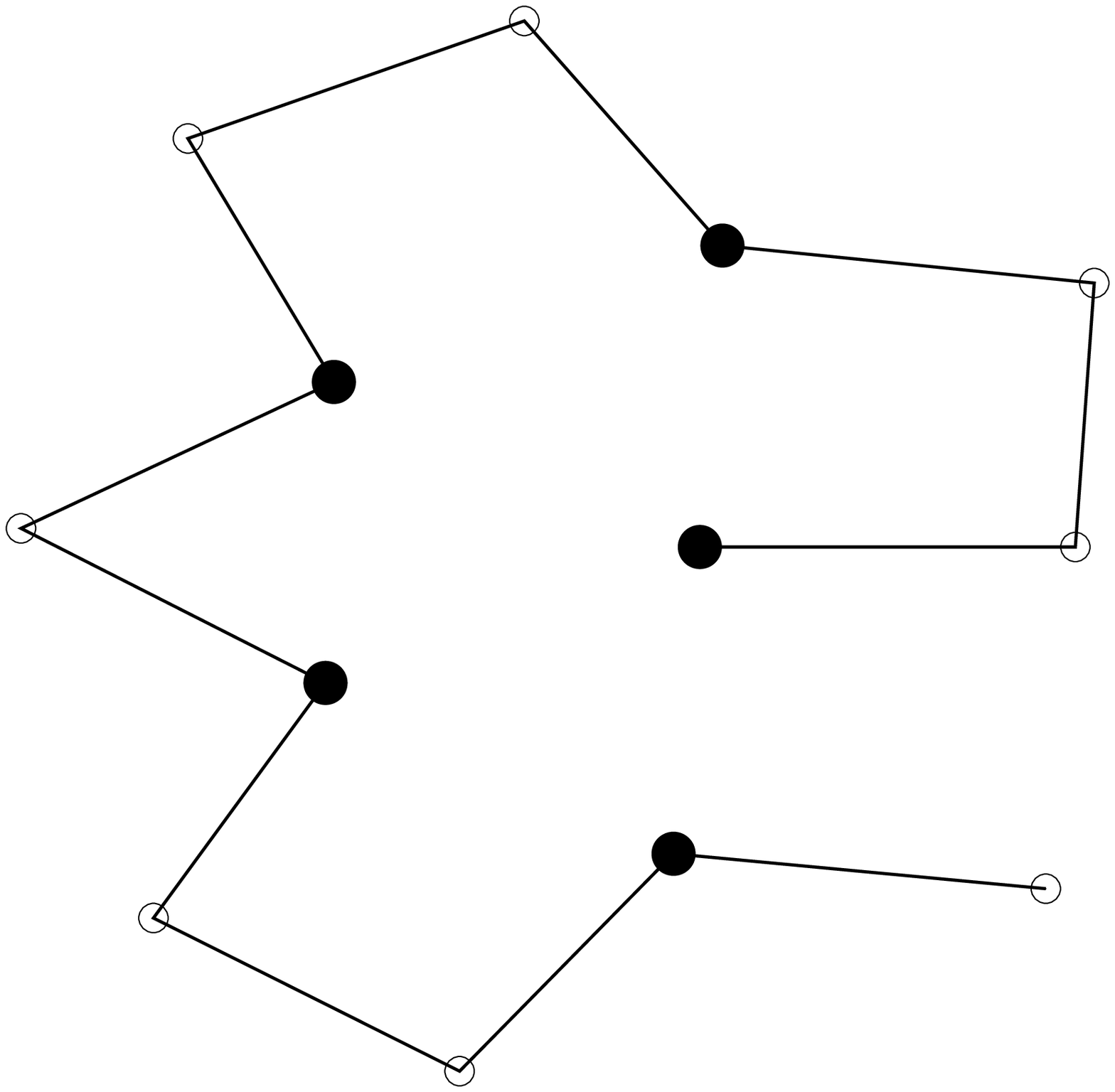, width=2.9cm, angle=270} &
\epsfig{file=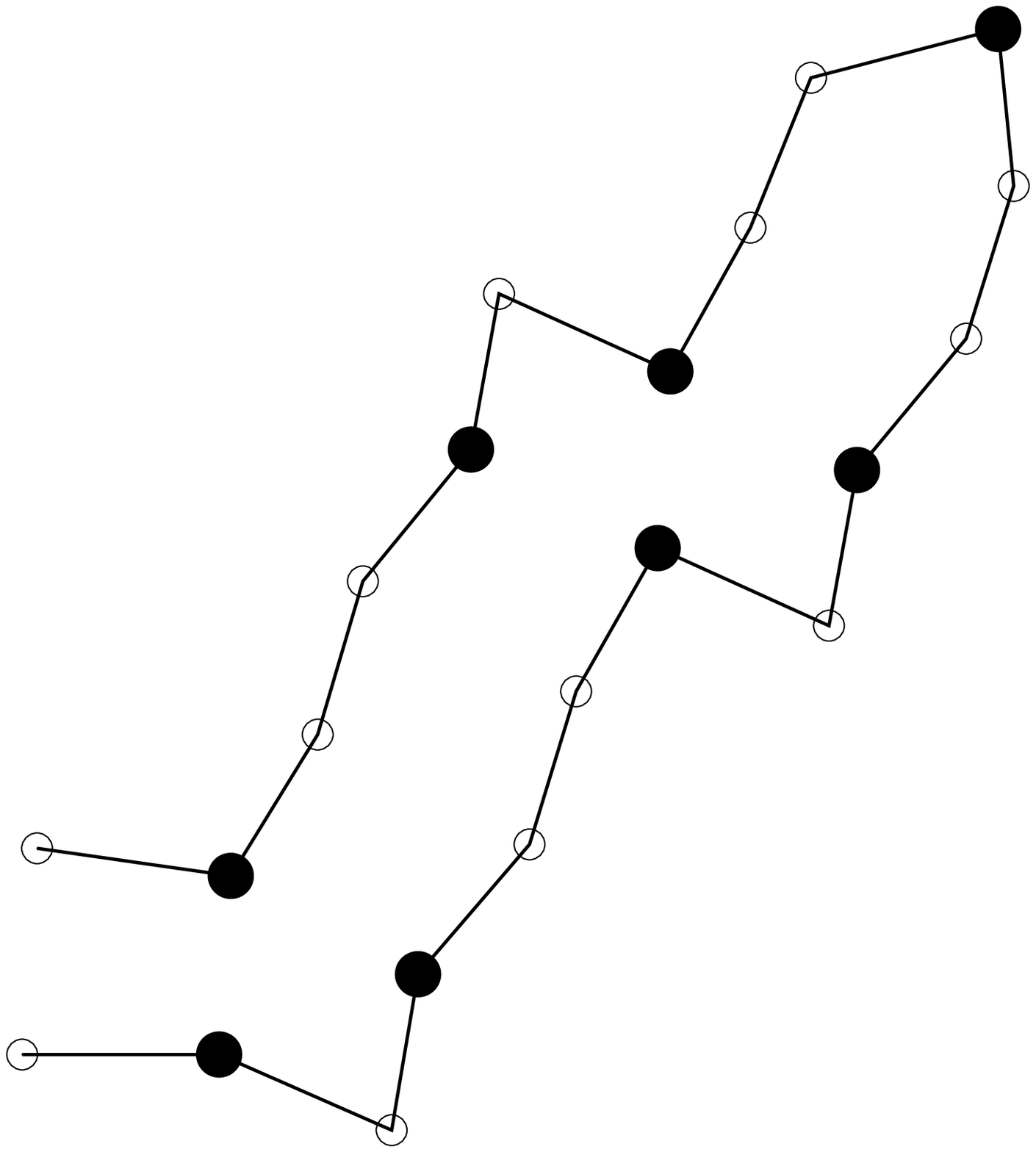, width=2.9cm, angle=270} \\
\multicolumn{1}{l}{\mbox{\tiny N = 34}} &
        \multicolumn{1}{l}{\mbox{\tiny N = 55}} \\ [-0.53cm]\\
\epsfig{file=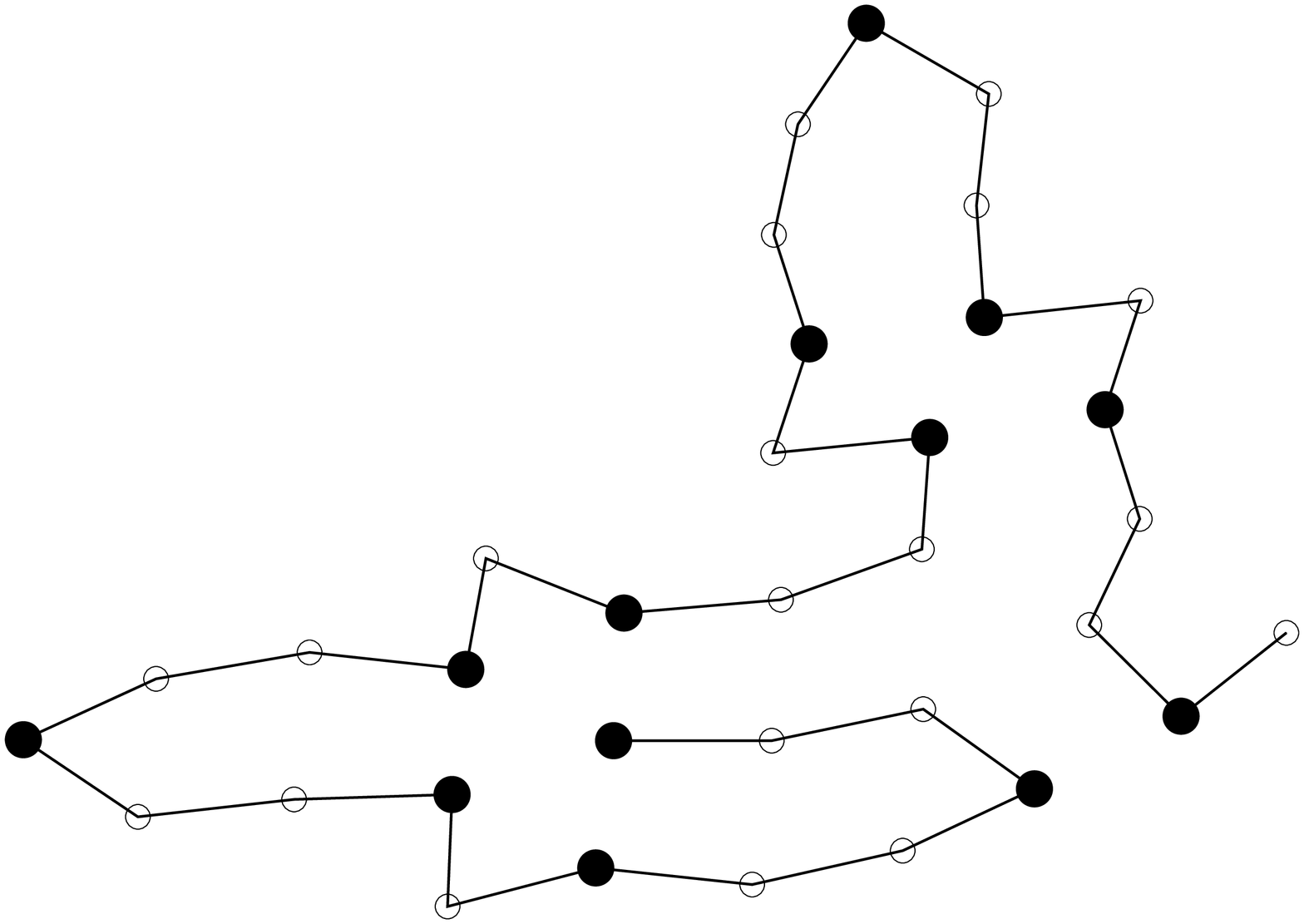, width=2.9cm, angle=270} &
\epsfig{file=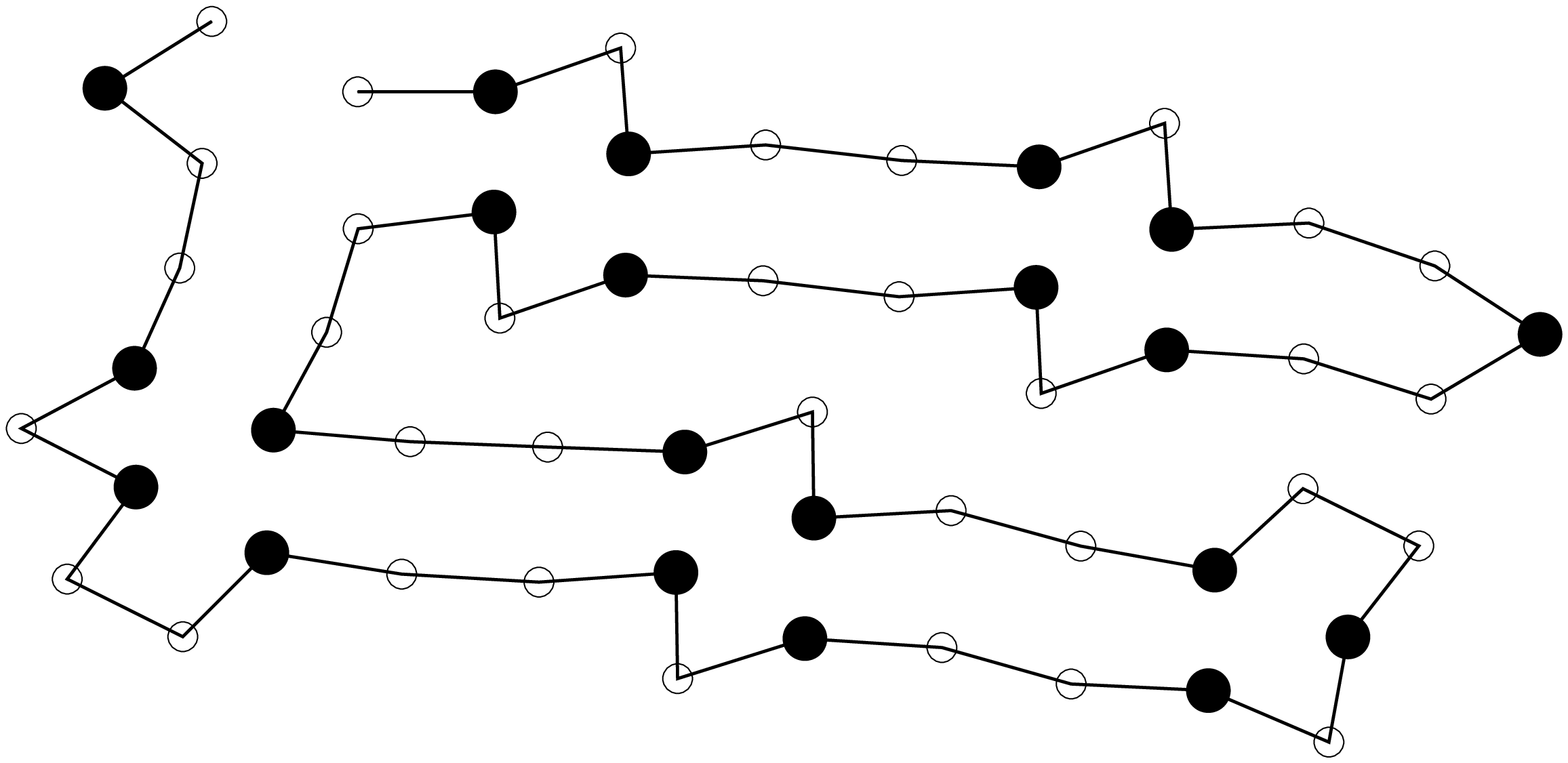, width=2.9cm, angle=270} \\
[0.4cm]
\end{array}$
\caption{ Putative ground states of Fibonacci sequences listed in Table I
in $2-d$ space. Full dots indicate hydrophobic monomers.}
\label{fig-2d}
\end{center}
\end{figure}

It is the purpose of the present paper to see whether PERM can be 
efficient for energy minimization in the AB model. In particular 
we shall use the new variant of PERM presented in \cite{hmng03}. 
We shall restrict ourselves to the subclass of ``Fibonacci sequences" 
studied also in \cite{sh95}, defined recursively by 
\be S_0=A,\quad S_1=B,\quad S_{i+1}=S_{i-1}*S_i \ee.
Here * is the concatenation operator. The first few sequences are 
$S_2 = AB,\;S_3 = BAB,\;S_4 = ABBAB$, etc. They have lengths given by $N_{i+1}=
N_{i-1}+N_i$, i.e given by the Fibonacci numbers. Hydrophobic residues ($A$)
occur isolated along the chain, while $B$s occur either isolated or in pairs.
The fraction of $B$s tends to the golden mean $\gamma=0.618~033$ as the 
length $N\rightarrow\infty$.

Although PERM gives also detailed information about excited states and 
thermodynamic behaviour at temperatures $T>0$, we shall not discuss this 
here. For studying the dynamics of the folding transition, in contrast, we 
would have to assume some realistic microscopic dynamics. Just like other 
advanced sampling methods such as simulated annealing or parallel tempering, 
PERM sacrifices the realism of the dynamics for efficiency. In addition,
as in the studies mentioned above, PERM is only used for coming close to 
the native state, and conjugate gradient descent is then used to reach the 
minimum energy state itself.

PERM is a biased chain growth algorithm with ``population control", i.e. a 
sequential importance sampling method with re-sampling \cite{liu}, implemented
recursively in a depth first fashion \cite{g97}. While chains grow, they acquire 
weights that include both Boltzmann factors and bias correction (``Rosenbluth"
\cite{rr55}) factors. During the growth, samples with large weight are cloned,
while chains with too small weight are pruned out. Except for the 
depth-first implementation and for the fact that it gives the correct 
Gibbs-Boltzmann statistics, PERM resembles therefore genetic algorithms.
While the original version of PERM was quite successful for lattice proteins
\cite{g98,bast98} and for a host of other applications \cite{perm-review}, 
it worked rather poorly for minimization of off-lattice polymer models \cite{fg99}.

In this paper we therefore employ an improved version called nPERMis in 
\cite{hmng03} (for "new PERM with importance sampling"). Basically, instead 
of making exact clones of high weight chains and hoping that these clones 
will evolve differently during the subsequent growth (as in original PERM), 
we now branch such that the last monomers are different at the point of 
branching. Thereby we now force the two copies to be distinct, and we 
avoid the loss of diversity that also plagues genetic algorithms when 
the evolution pressure is too high. 

\begin{figure}[h]
\begin{center}
$\begin{array}{c@{\hspace{0.1in}}c}
\multicolumn{1}{l}{\mbox{\tiny N = 13}} &
        \multicolumn{1}{l}{} \\ [-0.53cm]
\epsfig{file=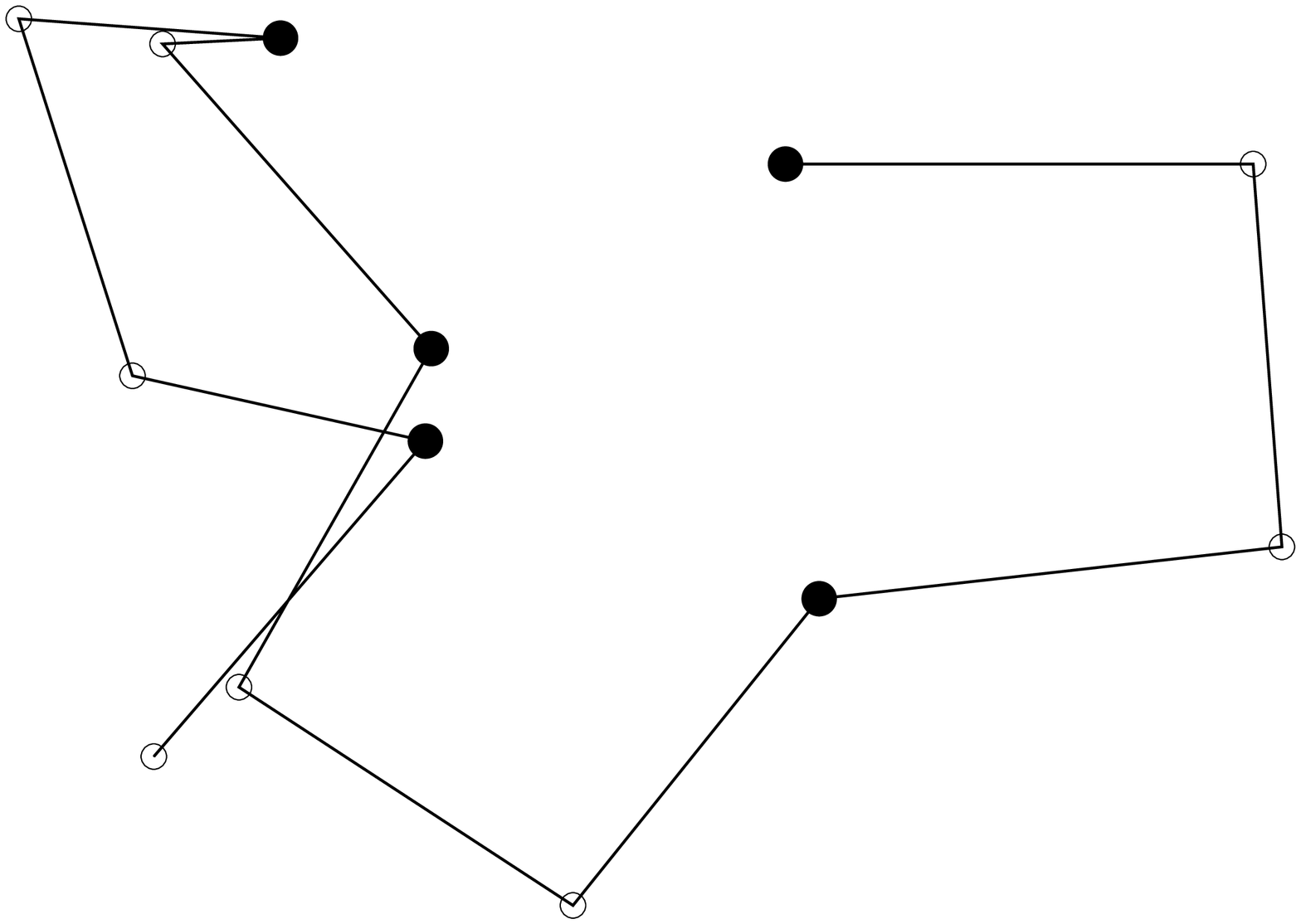, width=2.9cm, angle=270} &
\epsfig{file=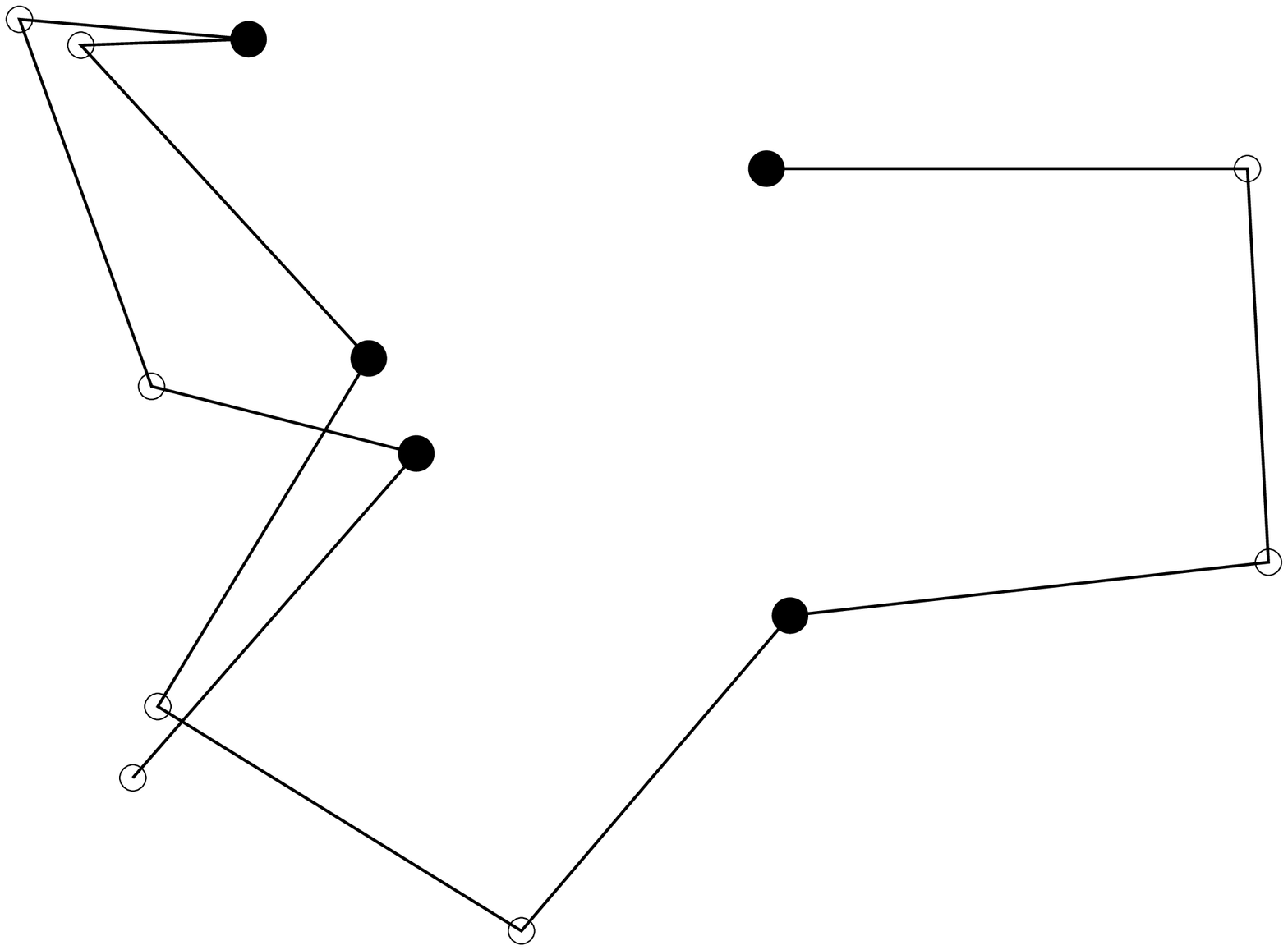, width=2.9cm, angle=270} \\
\multicolumn{1}{l}{\mbox{\tiny N = 21}} &
        \multicolumn{1}{l}{} \\ [-0.53cm]
\epsfig{file=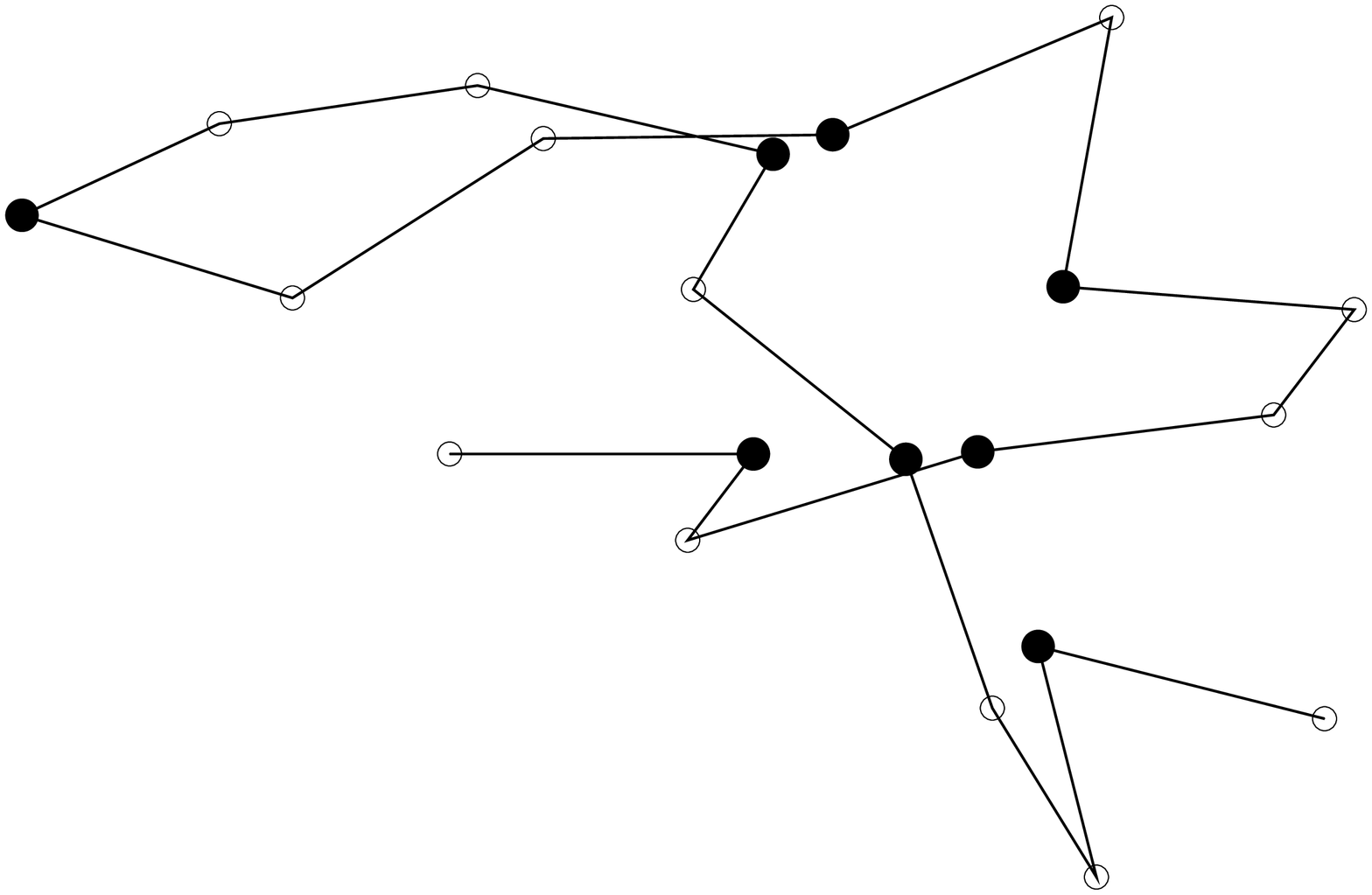, width=2.9cm, angle=270} &
\epsfig{file=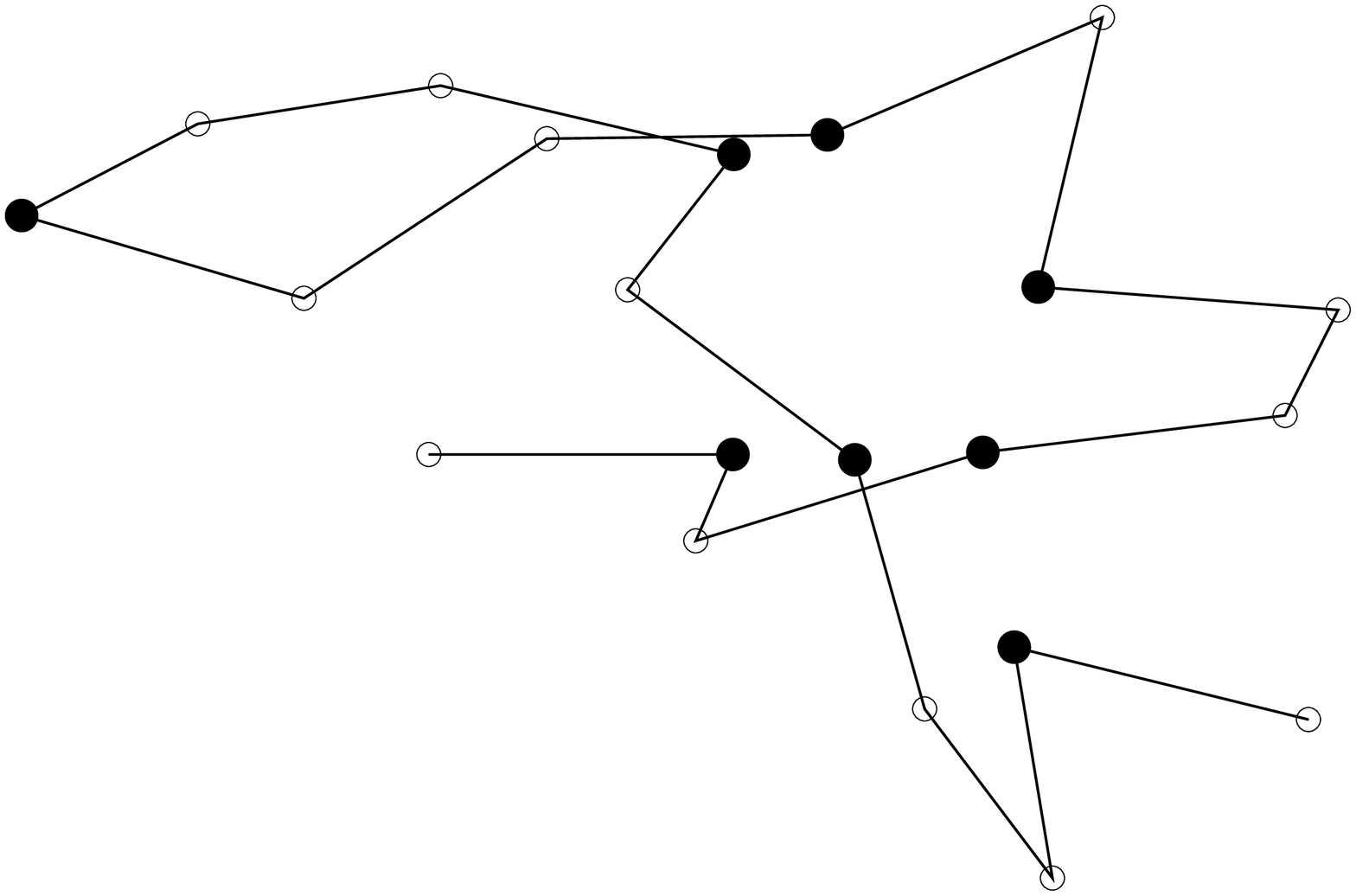, width=2.9cm, angle=270} \\
\multicolumn{1}{l}{\mbox{\tiny N = 34}} &
        \multicolumn{1}{l}{} \\ [-0.53cm]
\epsfig{file=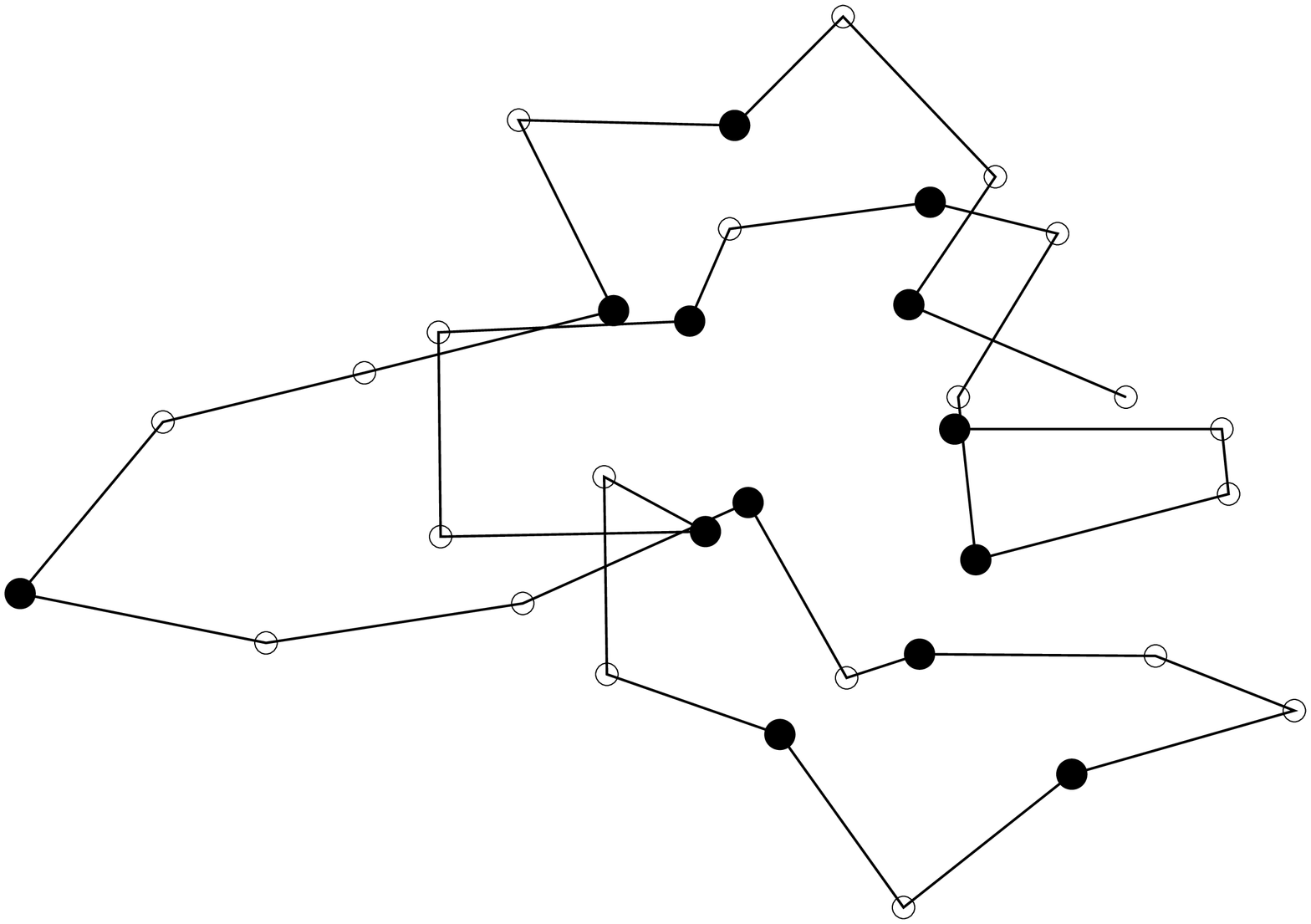, width=2.9cm, angle=270} &
\epsfig{file=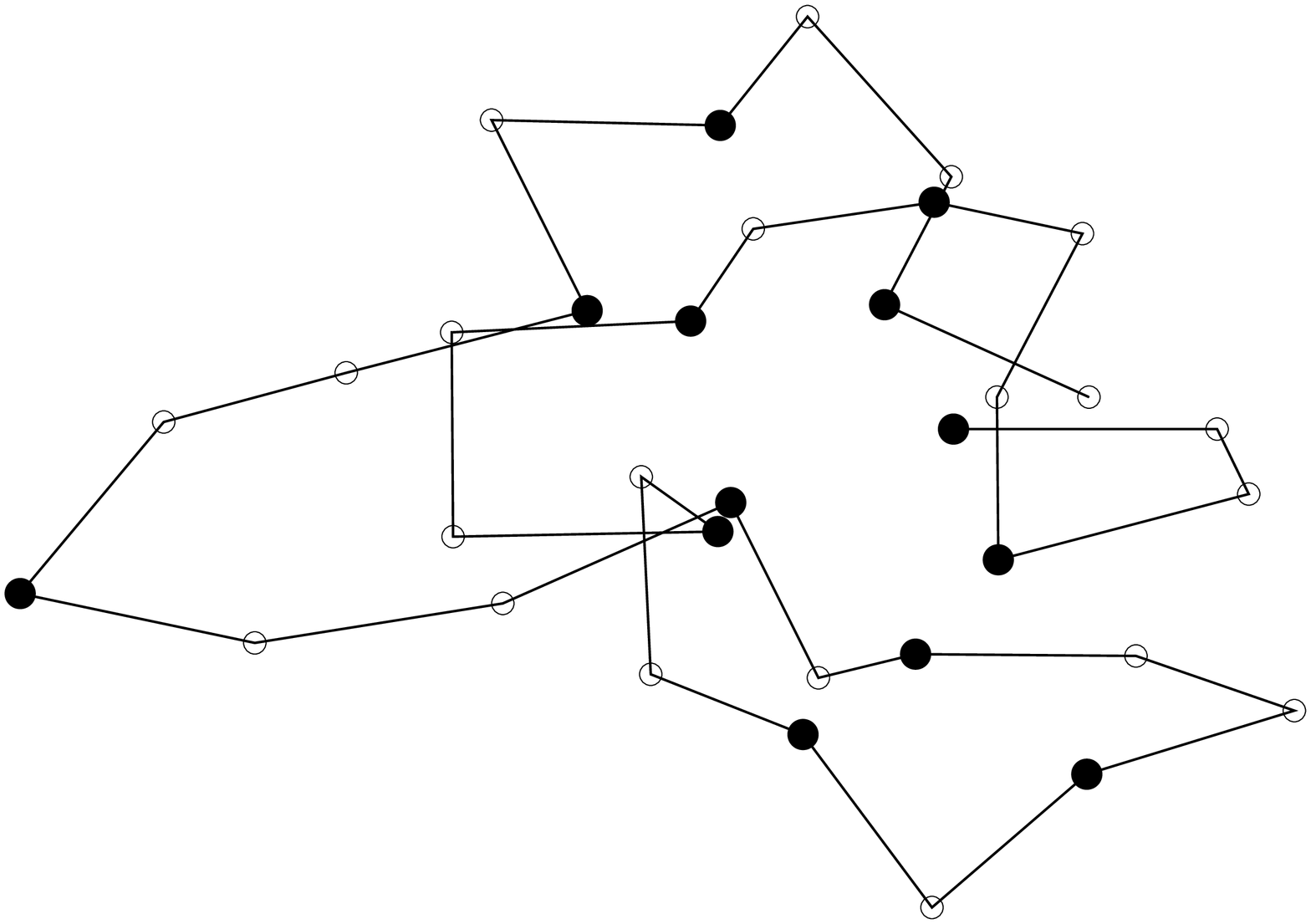, width=2.9cm, angle=270} \\
\multicolumn{1}{l}{\mbox{\tiny N = 55}} &
        \multicolumn{1}{l}{} \\ [-0.53cm]
\epsfig{file=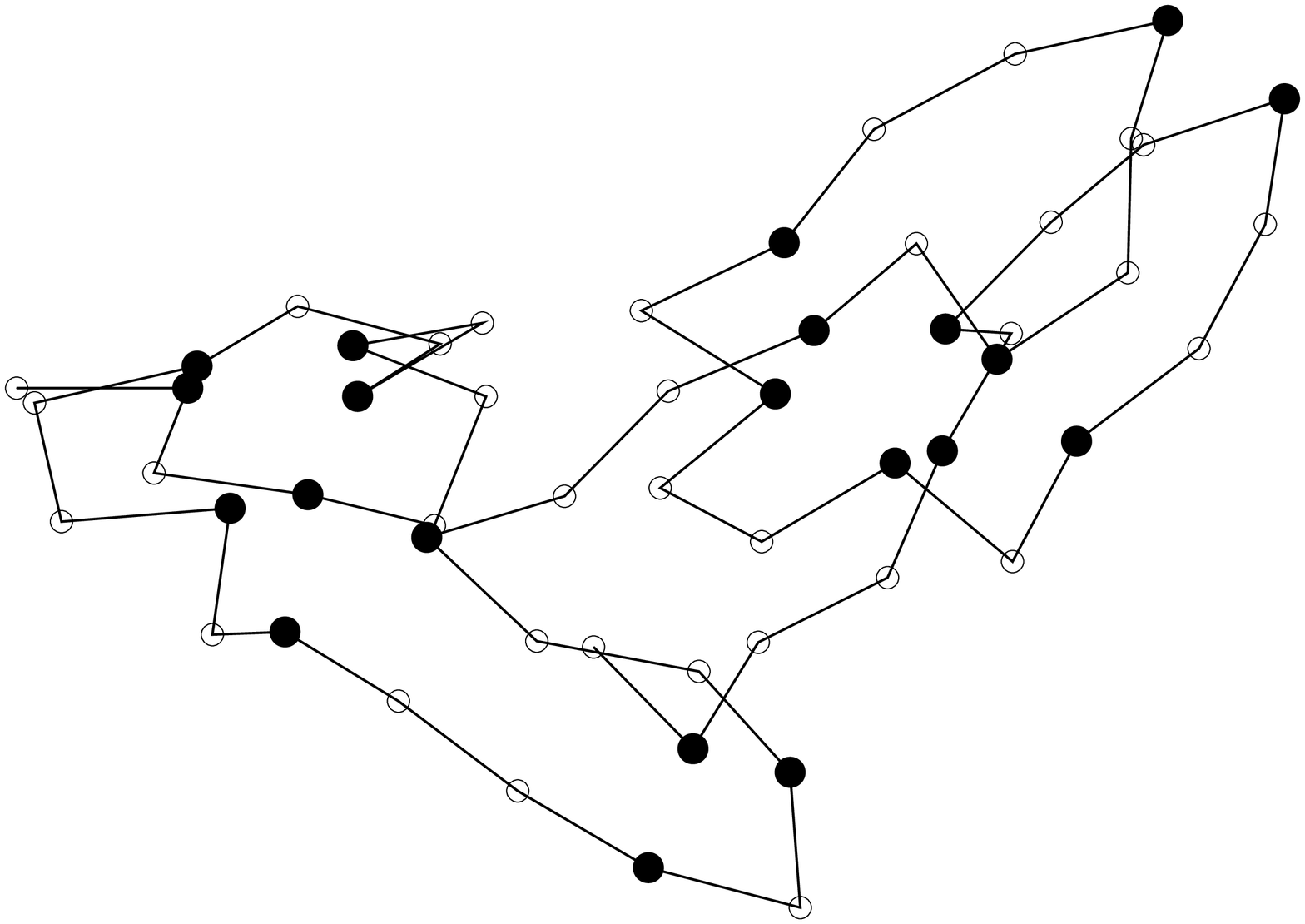, width=2.9cm, angle=270} &
\epsfig{file=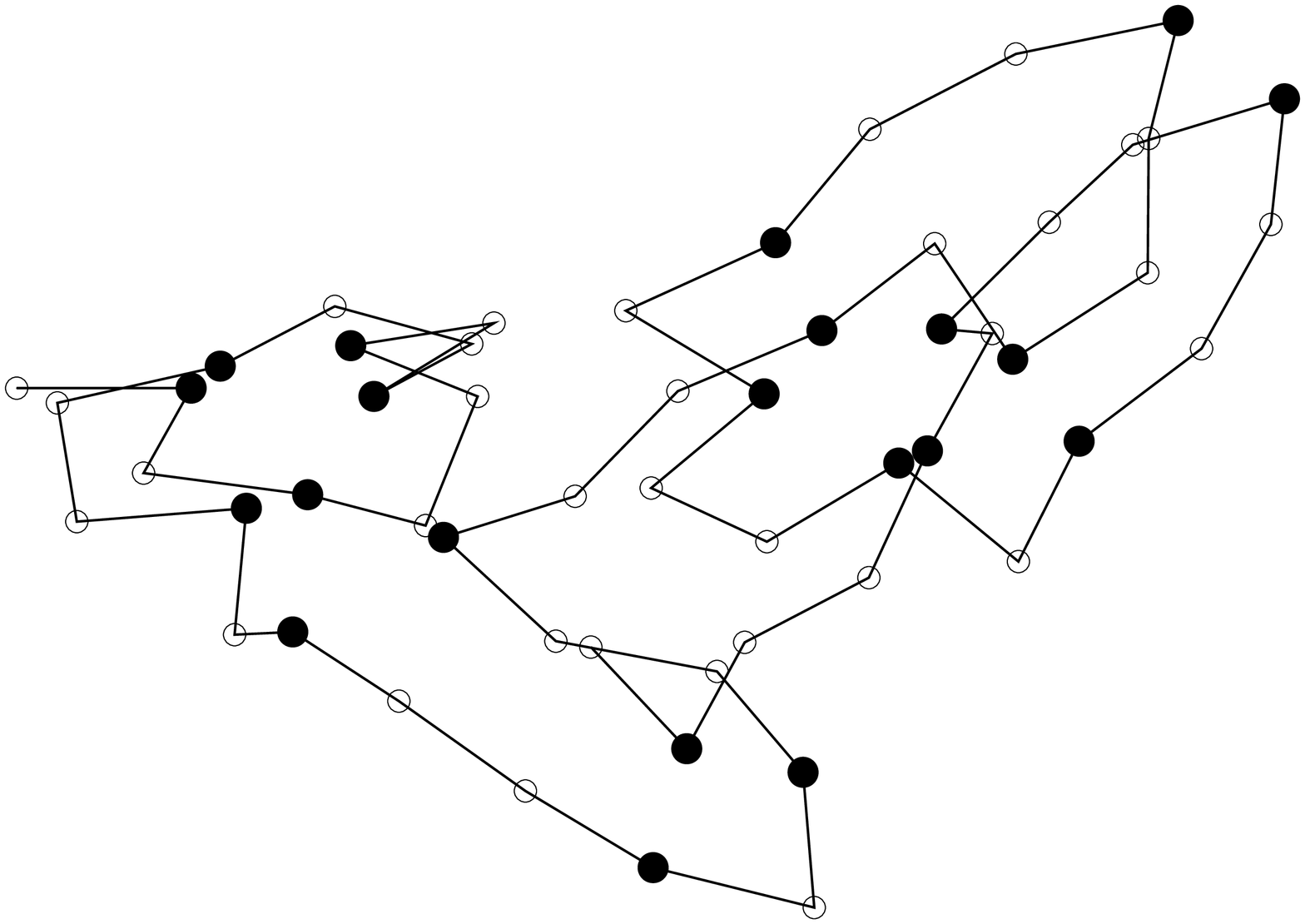, width=2.9cm, angle=270} \\
\end{array}$
\caption{ Stereographic views of putative ground states of $3-d$ Fibonacci
sequences listed in Table I. Again, $A$ monomers are shown as
filled circles.}
\label{fig-3d}
\end{center}
\end{figure}

\begin{table*}
\caption{Sequences and energies reached.
   $E_{\rm perm}$ is the lowest energy obtained by a PERM run, while
   $E_{\rm min}$ is the minimum energy obtained by subsequent conjugate gradient
   minimization. $E_{\rm min}^{*}$ is the putative ground state energy obtained
   by Stillinger and Head-Gordon\cite{sh95} (for $d=2$ only).}
\begin{ruledtabular}
\begin{tabular}{ccccccc} 
    &                         &    \multicolumn{3}{c}{$d=2$}        & \multicolumn{2}{c}{$d=3$} \\ 
$N$ & Sequence                & $E_{\rm perm}$ & $E_{\rm min}$ & $E_{\rm min}^*$ & $E_{\rm perm}$ & $E_{\rm min}$  \\ \hline
 13 & $ABBABBABABBAB$         &   -3.2167  &   -3.2939  &   -3.2235  &  -3.9730  &   -4.9616  \\
 21 & $BABABBABABBABBABABBAB$ &   -5.7501  &   -6.1976  &   -5.2881  &  -7.6857  &  -11.5238 \\                
 34 & $ABBABBABABBABBABABBAB$ &   -9.2195  &  -10.7001  &   -8.9749  & -12.8601  &  -21.5678\\
    & $ABBABBABABBAB$         &            &            &            &           &           \\
 55 & $BABABBABABBABBABABBAB$ &  -14.9050  &  -18.5154  &  -14.4089  & -20.1070  &  -32.8843\\
    & $ABBABBABABBABBABABBAB$ &            &            &            &           &          \\
    & $ABBABBABABBAB$         &            &            &            &           &          \\ 
\end{tabular}
\end{ruledtabular}
\end{table*}

For lattice polymers \cite{hmng03} one 
has for each partially grown chain a finite number of ``candidate directions"
for the next step. One first estimates the total weight of all these one-step
continuations. Based on this estimate, one decides on the number of clones
to be made. If, say, one wants to make $k$ clones, one scans all possible 
$k$-tuples of possible different candidate directions, and selects one of 
these tuples according to its weight. For off-lattice polymers one proceeds
exactly in the same way, with one exception: The candidates are now no longer 
the lattice bonds, but one has to choose $K$ candidate locations for the 
next monomer {\it randomly}. The number $K$ is an important parameter. While
$K\approx 5$ was optimal for 3-d polymers near the $\Theta$-point \cite{g97},
we found that lowest energies were reached in the AB model for $K \approx 50$.
While it was necessary to make clonings with very many siblings in simulating 
the HP model with the old version of PERM \cite{g98,bast98}, we now obtained
good results by restricting ourselves to $k$-tuples with $k\le 3$.

Another important parameter is the temperature at which the simulation is run. 
We typically used temperatures well below the collapse transition, $kT \approx 0.1$
or even lower. In order to speed up the ground state search, we also modified the 
Lennard-Jones potential by putting $E_2(r) = +\infty$ for $r<1$. This hard core
constraint
reduces the available phase space, but has no effect on ground state configurations
(we did not use it in the conjugate gradient minimization, and we checked that it
was satisfied after minimization). For chain deformation algorithms it could slow
down the dynamics, since the hard cores could act as barriers, but it can only
improve any pure chain growth algorithm. Finally, as a last trick, we used equally
spaced azimuthal angles for all candidates (with one overall angle chosen at random, 
for each group of candidates), in order to make them cover the unit sphere more 
uniformly. All simulations were done on Linux and UNIX workstations. CPU times were
up to 2 days, but their precise values are not very significant. Exact timings would
involve frequent comparisons of the minimizer basins of attraction reached by PERM, which 
we considered as too time consuming.

In Table I we list the lowest energies thus obtained for the two- and three-dimensional 
$AB$ model for all Fibonacci sequences with $13 \le N\le 55$. The latter is equal to 
the length of the longest sequence studied in \cite{sh95}. Let us first discuss the 
case $d=2$. For comparison we quote
also the putative ground state energies from Table II of \cite{sh95}. For $N< 13$, our 
energies agree perfectly with those of \cite{sh95}. Except for the shortest chain with 
$N=13$, already PERM gave in all cases shown in Table I lower energies than those found 
in \cite{sh95}. In all these cases already PERM by itself showed that the topologies
shown in \cite{sh95} are not the native ones. While the subsequent gradient descent 
improved the energies substantially, it changed in no case the overall topology.

The latter is true also for $d=3$, although there the subsequent minimization gave even 
larger energy changes than in $d=2$. This shows that in $d=3$, too, PERM is able to 
find states very close to the native ones. Since there exist no published ground state 
energies for the $3-d$ AB model, we are unable to compare PERM with other methods.

The configurations corresponding to the energies shown in Table I are shown in 
Figs.~1 (for $d=2$) and 2 (for $d=3$). For $d=2$ we see that none of the configurations, 
except the one for $N=13$, have single hydrophobic cores. Instead, the hydrophobic (A)
monomers form clusters of typically 4 to 5 particles. This is easily explained by 
the fact that hydrophobic monomers always are flanked by polar monomers along
the chain. Thus a clean separation into hydrophobic and polar regions is impossible.
This shows that the AB model with Fibonacci sequences would be a very poor model
for real proteins in $d=2$. From Fig.~3 we see that the same is true to a lesser 
degree in $d=3$. There the chains with $N=21$ and $N=34$ fold into configurations 
with single hydrophobic cores (except for a single A monomer which keeps out in both
cases), and only the chain with $N=55$ forms 2 clearly disjoint main hydrophobic 
groups.

In conclusion we have extended the PERM algorithm to an off-lattice 
two-species protein model. We have shown that it performs well, indeed we are 
able to refute with it previous claims for putative ground states. 

The chosen model is not very realistic. Partly this follows from the restriction
to two types of monomers, partly from the fact that we did not include, as in 
\cite{ipp3d}, more realistic local (bond angle and torsion) forces, and partly
from the restriction to Fibonacci sequences. Each of this features could have 
been easily avoided, and PERM works indeed equally well if we modify any of them.
But it was not our aim to present a realistic model. Rather we wanted to treat 
a model which is suitable for benchmarking, because of it is defined in a 
simple way and because it was already studied in detail before.  

It is less obvious whether PERM would also perform well for all-atom models with 
realistic potentials, or even with explicit solvents. Typically, its performance 
decreases quite rapidly with the number of degrees of freedom, but presumably 
it shares this with other modern methods like multicanonical sampling and 
parallel tempering. To answer this question, we have started such
simulations with the ECEPP force field implemented in SMMP \cite{smmp}. But
it is still too early to draw any conclusions.


Acknowledgments: We thank Walter Nadler for numerous very fruitful 
discussions and for critically reading the manuscript.


\begin{thebibliography}{30}
\bibitem{dill} K. F. Lau and K. A. Dill, Macromolecules {\bf 22}, 3986
   (1989); H. S. Chan and K. Dill, J. Chem. Phys. {\bf 95}, 3775 (1991);
   D. Shortle, H. S. Chan, and K. A. Dill, Protein Sci. {\bf 1}, 201 (1992).
\bibitem{yd93_95} K. Yue and K. A. Dill, Phys. Rev. E {\bf 48}, 2267 (1993);
   Proc. Natl. Acad. Sci. USA {\bf 92}, 146 (1995).
\bibitem{bd96} T. C. Beutler and K. A. Dill, Protein Sci. {\bf 5}, 2037 (1996).
\bibitem{chikenji} G. Chikenji, M. Kikuchi, and Y. Iba, Phys. Rev. Lett. {\bf
   83}, 1886 (1999); G. Chikenji and M. Kikuchi, Proc. Natl. Acad. Sci. USA
   {\bf 97}, 14273 (2000).
\bibitem{lw01} F. Liang and W. H. Wong, J. Chem Phys. {\bf 115}, 3374 (2001).
\bibitem{hmng03} H.-P. Hsu, V. Mehra, W. Nadler, and P. Grassberger, 
   J. Chem. Phys. {\bf 118}, 444 (2003); Phys. Rev. E {\bf 68}, 021113 (2003).
\bibitem{backofen} S. Will, "Constraint-based Hydrophobic Core Construction
for Protein Structure Prediction in the Face-Centered-Cubic
Lattice" in Proceedings of the Pacific Symposium on Biocomputing
2002 (PSB 2002) edited by Russ B. Altman et al.
(World Scientific, Singapore, 2002);
R. Backofen and S. Will, "Optimally Compact Finite Sphere
Packings --- Hydrophobic Cores in the FCC" in Proceedings
of the 12th Annual Symposium on Combinatorial Pattern
Matching (CPM 2001), Lecture Notes in Computer Sciences
2089, edited by Amihood Amir et al.
(Springer-Verlag, Berlin,  2001);
For a web application,
see http://www.bio.inf.uni-jena.de/Prediction/prediction.cgi
\bibitem{g97} P. Grassberger, Phys. Rev. E {\bf 56}, 3682 (1997).
\bibitem{shh93} F. H. Stillinger, T. Head-Gordon, and C. L. Hirshfeld, Phys. 
   Rev. E {\bf 48}, 1469 (1993); T. Head-Gordon and F. H. Stillinger Phys. Rev. 
   E {\bf 48}, 1502 (1993).
\bibitem{sh95} F. H. Stillinger and  T. Head-Gordon, Phys. Rev. E {\bf 52}, 2872 
   (1995). 
\bibitem{ipp3d}A. Irb\"{a}ck, C. Peterson, F. Potthast and O. Sommelius, J. Chem. 
   Phys. {\bf 107}, 273 (1997);
\bibitem{ipp}A. Irb\"{a}ck, C. Peterson and F. Potthast, Phys. Rev. E {\bf 55}, 
860 (1997); A. Irb\"{a}ck and F. Potthast, J. Chem Phys. {\bf 103}, 10298 
(1995).
\bibitem{gorse02} D. Gorse, Biopolymers {\bf 64}, 146 (2002).
\bibitem{tlp01} A. Torcini, R. Livi, and A. Politi, J. Bio. Phys. {\bf 27}, 181 
(2001). 
\bibitem{gorse01} D. Gorse, Biopolymers {\bf 59}, 411 (2001).
\bibitem{liu} J.S. Liu, {\it Monte Carlo Strategies in Scientific Computing},
    Springer Series in Statistics (Springer, New York 2001)
\bibitem{rr55} M. N. Rosenbluth and A. W. Rosenbluth, J. Chem. Phys. {\bf 23},
   356 (1955).
\bibitem{g98} H. Frauenkron, U. Bastolla, E. Gerstner, P. Grassberger, and W.
   Nadler, Phys. Rev. Lett. {\bf 80}, 3149 (1998).
\bibitem{bast98} U. Bastolla, H. Frauenkron, E. Gerstner, P. Grassberger, and W.
   Nadler, Proteins {\bf 32}, 52 (1998).
\bibitem{perm-review} P. Grassberger, Comp. Phys. Comm. {\bf 147}, 64 (2002).
\bibitem{fg99} H. Frauenkron and P. Grassberger, unpublished.
\bibitem{smmp} F. Eisenmenger, U. H. E. Hansmann, Sh. Hayryan and C.-K. Hu, Comp. Phys. Comm., {\bf 138}, 192 (2001). 

\end{thebibliography}
\end{document}